\documentclass[reprint,amsmath,amssymb,aps,pra
]{revtex4-2}

\usepackage{graphicx}
\usepackage{dcolumn}
\usepackage{bm}
\usepackage[utf8]{inputenc}
\usepackage{nicefrac}
\usepackage{natbib}
\usepackage{amsmath}
\usepackage[dvipsnames]{xcolor}
\usepackage[T1]{fontenc}
\usepackage{mathptmx}
\usepackage{etoolbox}
\usepackage{booktabs}
\usepackage{hyperref}
\usepackage{soul}
\newcommand{\RNum}[1]{\uppercase\expandafter{\romannumeral #1\relax}}
\usepackage{xcolor}
\usepackage[dvipsnames]{xcolor}

\makeatletter

\def\@email#1#2{%
 \endgroup
 \patchcmd{\titleblock@produce}
  {\frontmatter@RRAPformat}
  {\frontmatter@RRAPformat{\produce@RRAP{*#1\href{mailto:#2}{#2}}}\frontmatter@RRAPformat}
  {}{}
}%
\makeatother

\begin{document}
\preprint{APS/123-QED}

\title{Laser spectroscopy on the hyperfine structure and isotope shift of\\sympathetically cooled $^{229}$Th$^{3+}$ ions}

\author{G. Zitzer$^1$}
\author{J. Tiedau$^1$} 
\author{Ch. E. Düllmann$^{2,3,4}$}
\author{M. V. Okhapkin$^1$}
\author{E. Peik$^{1}$}
\email{email: ekkehard.peik@ptb.de}
\affiliation{$^1$ Physikalisch-Technische Bundesanstalt, Braunschweig, Germany}
\affiliation{$^2$ Johannes Gutenberg University Mainz, Mainz, Germany}
\affiliation{$^3$ Helmholtz Institute Mainz, Mainz, Germany}
\affiliation{$^4$ GSI Helmholtzzentrum für Schwerionenforschung GmbH, Darmstadt, Germany}
\vspace{10pt}
\date{April 1, 2025}

\begin{abstract}
The hyperfine structure of $^{229}$Th$^{3+}$ ions in the nuclear ground state is investigated via laser spectroscopy of trapped Th$^{3+}$ ions that are sympathetically cooled by laser-cooled $^{88}$Sr$^+$ ions in a linear Paul trap. The isotope shift to $^{230}$Th$^{3+}$ and the hyperfine constants for the magnetic dipole (\textit{A}) and electric quadrupole (\textit{B}) interactions for the 5F$_{5/2}$ and 6D$_{5/2}$ electronic states of $^{229}$Th$^{3+}$ are determined. These measurements provide nuclear moments of $^{229}$Th with reduced uncertainty and serve as a preparation for improved hyperfine spectroscopy of the 8.4~eV nuclear isomeric state in $^{229}$Th$^{3+}$ ions.

\end{abstract}

\maketitle

The isotope $^{229}$Th is of particular interest because of its low-energy 8.4-eV-isomer. Recently, three experiments have succeeded in driving the magnetic dipole transition at 148\,nm wavelength from the nuclear ground state to the isomer in $^{229}$Th-doped crystals with tunable lasers ~\cite{Tiedau:2024,ElwellHudson:2024,Zhang:2024}. Among several research topics at the interface of nuclear and atomic physics, the most promising application of this system seems to be a highly precise optical nuclear clock ~\cite{Peik:2003,Campbell:2012,Peik:2021}. While the study of $^{229}$Th with precision techniques of laser nuclear spectroscopy has just begun, the excitation energy for $^{229}$Th in calcium fluoride crystals has already been reported with a relative uncertainty in the range of $10^{-12}$ ~\cite{Zhang:2024}. For an assessment of the achievable performance of a $^{229}$Th-based nuclear clock and for an improved understanding of the nuclear structural changes underlying the low-energy transition, knowledge of the nuclear magnetic moments and of the charge distribution of the ground state and isomer is required. A few studies of the hyperfine structure of electronic transitions in the $^{229}$Th ground state have been performed in the past ~\cite{Gerste:1974, Kaelber:1989}. The highest accuracy has been obtained so far in an experiment with laser-cooled, trapped $^{229}$Th$^{3+}$ ions ~\cite{Campbell:2011}, combined with \textit{ab initio} atomic structure calculations~\cite{Safronova:2013,Porsev:2021}. 

The hyperfine structure of electronic transitions of the 8.4-eV-isomer $^{229m}$Th has been observed so far in only two experiments~\cite{Thielking:2018,Yamaguchi:2024}, where hyperfine splittings  have been compared for $^{229}$Th and $^{229m}$Th. In this way, the nuclear moments of the isomer have been determined with reference to the ground state values from Ref.~\cite{Campbell:2011,Safronova:2013,Porsev:2021}. In agreement with the expectations from a liquid-drop model, the intrinsic electric quadrupole moments of $^{229m}$Th and $^{229}$Th have been found to be equal to within the uncertainty, but the experimental result for the magnetic moment of $^{229m}$Th differed significantly from theory ~\cite{Dykhne:1998matrix}. 
The recent measurement of the quadrupole splitting of the nuclear resonance in a Th:CaF$_2$ crystal, in the presence of the strong gradient of the crystal electric field, has provided the value for the ratio of the $^{229m}$Th\,/\,$^{229}$Th intrinsic electric quadrupole moments: 1.01791(2) ~\cite{Beeks:2024_finestructureconstant}. 
Recent nuclear structure calculations reproduce the near-equality of the electric quadrupole moments, but remain with a substantial uncertainty on the magnetic moments ~\cite{Minkov:2024}. With the aim to strengthen the experimental database on this topic, the first improved, independent measurement since Campbell \textit{et al.} (Ref.~\cite{Campbell:2011}), of the Th-229 nuclear moments from laser spectroscopy of trapped, sympathetically cooled $^{229}$Th$^{3+}$ ions, and additionally the isotope shift to $^{230}$Th$^{3+}$, is presented here. Since the recoil ion source also provides a 2\,\% fraction of $^{229}$Th ions in the isomeric state \cite{Barci:2003,Thielking:2018}, an experiment for measurements of the hyperfine structure with sympathetically cooled isomer ions is planned with the setup.

The experimental setup consists of a beamline for the generation and guiding of thorium ions, a linear Paul trap, and an optical setup. The ion beamline is used for loading Th$^{3+}$ recoil ions from the $\alpha$-decay of a uranium source into the ion trap. The optical setup consists of lasers for cooling of Sr$^{+}$ and for spectroscopy of Th$^{3+}$ ions. Detailed descriptions of the beamline, the linear radio-frequency ion trap, and the optical system for the sympathetic cooling of Th$^{3+}$ are presented in Ref.~\cite{Zitzer:2024}. The laser setup described in~\cite{Zitzer:2024} is partially modified for precision measurements of the hyperfine structure (hfs) in $^{229}$Th$^{3+}$.

Two uranium sources, $^{233}$U and $^{234}$U, produced by molecular plating ~\cite{Haas:2020}, are implemented in the system so that either $^{229}$Th$^{3+}$ or $^{230}$Th$^{3+}$ ions are produced for the experiments. For selecting the isotope that enters the beamline, the two sources are mounted back to back on a vacuum rotary feedthrough. After leaving the source region filled with helium buffer gas for dissipation of the recoil kinetic energy, the Th$^{3+}$ ions are transported and collected in a rf quadrupole (RFQ). From there, the ions are extracted in bunches into a quadrupole mass filter (QMF) and are redirected by a 90° deflector towards the ion trap. An octupole ion guide and a trap focusing lens are used to inject the ions into the trap where the sympathetic cooling of Th$^{3+}$ with Sr$^{+}$ ions is performed in a dedicated segment of the ion trap ~\cite{Zitzer:2024}. An additional end cap electrode was added to the ion trap to improve the trapping efficiency and confinement for the thorium ions. Differential pumping is used to reduce the helium pressure from $10^3$~Pa in the source region to $10^{-7}$~Pa in the ion trap chamber. 

\begin{figure}[ht]
\includegraphics[scale=0.52]{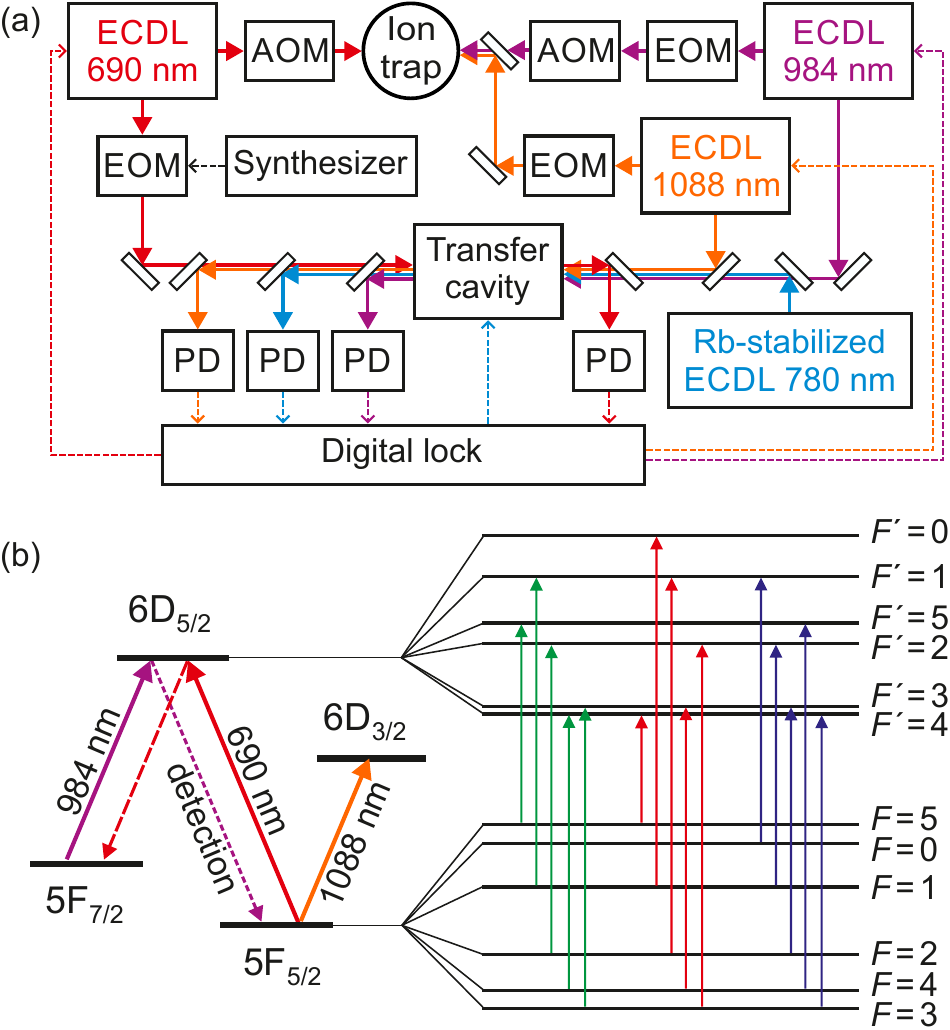}
\caption{\label{fig:setup} Optical setup (a) for the spectroscopy of sympathetically cooled $^{229}$Th$^{3+}$ ions and the level scheme (b) with relevant hyperfine transitions (see Ref.~\cite{Campbell:2011}). The ECDL at 690\,nm is used for the excitation of thorium ions. To provide precise tuning of this laser, a fraction of its radiation is modulated via an EOM and a modulation sideband is locked to the transfer cavity. The radiation of an ECDL at 984\,nm is broadened to about 3\,GHz via a fiber-coupled broadband EOM for the depopulation of the 5F$_{7/2}$ state. Similar broadening is performed via the EOM of the 1088\,nm ECDL.
 The 984\,nm and 1088\,nm laser are frequency locked to the transfer cavity which length is stabilized using the radiation of a Rb-stabilized ECDL. The transfer cavity transmission signals of the spectroscopy lasers are separated and detected by photodiodes (PD). The digital lock stabilizes the positions of the cavity resonances relatively to the Rb-stabilized laser.} 
\end{figure}

The excitation scheme and the laser setup used in the experiment are shown in Fig.~\ref{fig:setup}. 
For the spectroscopy of the $^{229}$Th$^{3+}$ 5F$_{5/2}$\,$\rightarrow$\,6D$_{5/2}$ transition an external cavity diode laser (ECDL) at 690\,nm is used. A small fraction of the laser radiation passes a broadband fiber-coupled electro-optical modulator (EOM). The frequency of one of the sidebands formed by the EOM is locked to a transfer cavity, see Fig.~\ref{fig:setup}(a). By varying the frequency of a synthesizer that drives the EOM the carrier frequency of the laser radiation can be tuned widely with high precision. The unmodulated fraction of the ECDL radiation at 690\,nm is used for the excitation of the ions. 
For the detection of the transition 5F$_{5/2}$\,$\rightarrow$\,6D$_{5/2}$ a $\approx$\,100\,µs excitation pulse of the 690\,nm laser is applied, formed by an acousto-optic modulator (AOM). This transfers population to the 5F$_{7/2}$ level through the spontaneous decay from the 6D$_{5/2}$ state~\cite{Zitzer:2024}.
Next, a pulse of the 984\,nm radiation is applied for about 100\,µs and the fluorescence signal at 690\,nm from the 6D$_{5/2}$ level decay to the ground state is registered. This scheme eliminates the appearance of 690\,nm laser stray light in the fluorescence detection. The frequency of the 984\,nm laser is modulated by another fiber-coupled EOM to depopulate all of the hyperfine sub-levels of the 5F$_{7/2}$ state.
For the measurements on sympathetically cooled $^{229}$Th$^{3+}$ ions, laser powers in the range of 30\,µW - 180\,µW for the transition at 690\,nm are used. The laser power for the transition at 984\,nm is set to about 800\,µW. The beam diameters of the 690\,nm and 984\,nm lasers
are $\approx$\,0.5\,mm in the trap region. In addition, laser light at 1088\,nm from an ECDL with a power of $\approx 15$\,mW, modulated to a width of about 3 GHz via an EOM, is applied continuously to mix the population of the ground state hyperfine sublevels. 
The fluorescence of the transition at 690\,nm is detected using a photomultiplier tube with a corresponding bandpass interference filter, as also described in Ref.~\cite{Zitzer:2024}.

The laser spectroscopy measurements are performed with $^{229}$Th$^{3+}$ ions that are embedded and sympathetically cooled by Coulomb crystals of Sr$^{+}$ ions, similar to the ones shown in Ref.~\cite{Zitzer:2024} and consisting of typically a few hundred Sr$^+$ and tens of Th$^{3+}$ ions. 
From the previous experiments on $^{230}$Th$^{3+}$~\cite{Zitzer:2024}, an upper limit of 140\,mK is deduced for the temperature of the trapped ions.
Because of the low Th$^{3+}$ ion number and recording times that exceeded the isomer lifetime, a fluorescence signal from the isomer is not observed here.

The hfs of $^{229}$Th$^{3+}$ in the nuclear ground state is investigated for the 5F$_{5/2}$\,$\rightarrow$\,6D$_{5/2}$ (690\,nm) transition. The nuclear spin is $I\,=\,5/2$. Taking into account the angular momentum $\vec{J}$ of the atomic shell and the nuclear spin $\vec{I}$, with $\vec{F} = \vec{J} + \vec{I}$, each state splits into 6 hfs sub-levels with a total number of 15 transitions according to the selection rules $\Delta F = 0, \pm 1$; see Fig.~\ref{fig:setup}(b).
The associated energy, with the hfs constants \textit{A} (magnetic dipole) and \textit{B} (electric quadrupole), neglecting higher orders by e.g. octupole contributions, is calculated as follows \cite{Schwartz:1955}: 

\begin{equation}\label{eqn:hfsenergie}
E_{hfs} = A \frac{M(I,J,F)}{2} + B \frac{N(I,J,F)}{4}
\end{equation}
\begin{align*}
\text{with}&& M(I,J,F) =&& F(F+1) - I(I+1) - J(J+1)&\\
\text{and} && N(I,J,F) =&& \frac{3 M (M + 1) - 4I(I+1) J(J+1)}{2I(2I-1) J(2J-1)}
\end{align*}

\begin{figure}[t]
\includegraphics[scale=0.3]{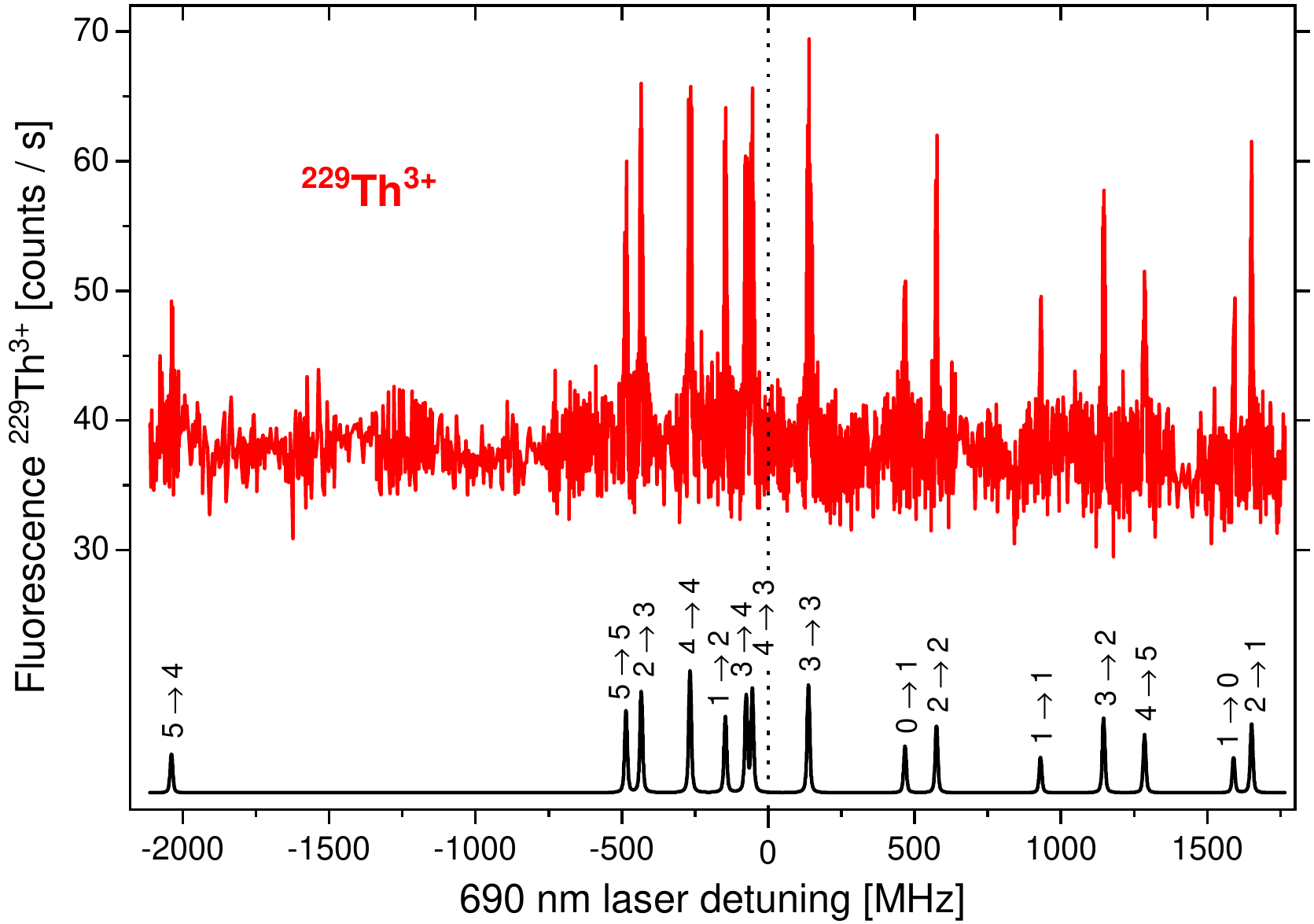}
\caption{\label{fig:HFSspektrum} Observed hyperfine spectrum of the 5F$_{5/2}$\,$\rightarrow$\,6D$_{5/2}$ transition of $^{229}$Th$^{3+}$. The lower part of the figure shows a scaled fit and the corresponding quantum numbers \textit{F,\,F'} of the individual hfs components. A part of the spectrum is illustrated in more detail in Fig.~\ref{fig:Isotopenshiftbild}.}
\end{figure}

In Fig. \ref{fig:HFSspektrum} an excitation spectrum of the performed measurements of the 690\,nm transition is presented. All 15 hfs components of the 5F$_{5/2}$\,$\rightarrow$\,6D$_{5/2}$ transition are resolved. Below the experimental data, a scaled fit is shown, also indicating the corresponding quantum numbers \textit{F} for each transition. 

The linewidth of the observed resonances is predominantly determined by power broadening which was investigated previously in Ref.~\cite{Zitzer:2024}. The power broadening of the resonances is analyzed for the $^{229}$Th$^{3+}$ hfs transition $\vert$5F$_{5/2}$, \textit{F}\,=\,2$\rangle \rightarrow \vert$6D$_{5/2}$,\,$F'=\,2\rangle$ to find the excitation parameters. For the lowest laser powers of approximately 30\,µW at wavelength 690\,nm, this results in a full width at half maximum (FWHM) of about 10 MHz. The laser power dependence is similar to the previous measurements with $^{230}$Th$^{3+}$. 

\begin{table*}[!]
\renewcommand{\arraystretch}{1.35}
\begin{tabular}{|c |cc|cccc|}
\hline
\multicolumn{1}{|l|}{} & \multicolumn{2}{c|}{\textbf{Ref.~\cite{Campbell:2011}}} & \multicolumn{4}{c|}{\textbf{This work}} \\[0.2 ex] 
State & \textit{A} {[}MHz{]} & \textit{B} {[}MHz{]} & \textit{A} {[}MHz{]} & \multicolumn{1}{c|}{\textit{B} {[}MHz{]}} & $\mu/\mu_N$ & $Q/e~[10^{-28}\mathrm{\,m}^2]$  \\ \hline
5F$_{5/2}$ & 82.2(6) & 2269(6) & 82.0(2) &
 \multicolumn{1}{c|}{2270.3(1.8)} &  0.365(3) &  3.11(5) \\
6D$_{5/2}$ &  -12.6(7) &
  2694(7) &
  -12.9(3) &
  \multicolumn{1}{c|}{2695.7(1.9)} & 0.31(7) &
  3.10(4) \\ \hline
\end{tabular}

\caption{Hyperfine constants \textit{A} and \textit{B} of the 5F$_{5/2}$\,$\rightarrow$\,6D$_{5/2}$ transition at 690\,nm for $^{229}$Th$^{3+}$ from a fit to the measured spectra. The previous values from Ref.~\cite{Campbell:2011} are listed for comparison. The nuclear moments, $\mu$ and $Q$, are calculated using the theoretical atomic structure coefficients from Ref.~\cite{Porsev:2021}.}
\label{hfsAB-alt}
\end{table*}

The measured hfs spectra are analyzed using the fitting software \textit{SATLAS2} \cite{Sat:2024}. The software uses Voigt profiles for the lineshapes of the individual peaks. Racah coefficients, which fix the relative peak intensities, are considered as initial fit parameters. For final fitting, the peak heights are switched to free parameters to match the recorded signals. The resulting hyperfine constants \textit{A} and \textit{B} for the 5F$_{5/2}$ and for the 6D$_{5/2}$ states are presented in Table \ref{hfsAB-alt}. The uncertainties contain systematic and statistical contributions of similar size. The dominant contribution to the systematic uncertainties originates from the instability of the transfer cavity, which is used for the tuning and stabilization of the laser frequencies.

\begin{table*}[ht]
\renewcommand{\arraystretch}{1.35}
\begin{tabular}{|l|c|c|c|}
\hline
 & $\delta\nu^{229,230}$ {[}MHz{]} & $\delta\nu^{229,232}$ {[}MHz{]} & $\delta\nu^{229,232} / \delta\nu^{229,230}$ \\ \hline
Th$^{+}$ (584 nm), Ref.~\cite{Kaelber:1989} & 9650(95) & 25010(90) & 2.592(28) \\
Th$^{3+}$ (690 nm), this work$^{\dagger}$ and Ref.~\cite{Campbell:2011}$^{\ddagger}$ & -4055.9(2.4)$^{\dagger}$ & -10509(7)$^{\ddagger}$ & 2.591(3) \\ \hline
\multicolumn{1}{|c|}{$\delta\nu$(Th$^{+}$) / $\delta\nu$(Th$^{3+})$} & 2.379(24) & 2.380(9) & \\ \hline
\end{tabular}
\caption{Compilation of precision measurements of isotope shifts of two thorium resonance lines: in Th$^{+}$ at 583.9\,nm, from Ref.~\cite{Kaelber:1989}, and in Th$^{3+}$ at 690\,nm, from this work and Ref.~\cite{Campbell:2011}. Similar to a King plot analysis, the ratio of the isotope shifts of the two transitions $\delta\nu$(Th$^{+}$) / $\delta\nu$(Th$^{3+})$ can be interpreted as the ratio of the field shift coefficients for the two transitions. The agreement of the ratio $\delta\nu^{229,232} / \delta\nu^{229,230}$ of the isotope shifts for the two isotope pairs for the charge states $1+$ and $3+$ indicates the consistency of the spectroscopic measurements.}
\label{isotopeshift}
\end{table*}

The measured hyperfine constants \textit{A} and \textit{B} agree with the values obtained by Campbell \textit{et al.} \cite{Campbell:2011} within the uncertainties. With the measurements presented here, the uncertainties on the \textit{A} and \textit{B} factors are reduced by up to a factor of about three (see Tab.~\ref{hfsAB-alt}).

The hfs constants \textit{A} and \textit{B} are linked to the magnetic dipole moment \textit{µ} and the electric quadrupole moment \textit{Q} of the $^{229}$Th nucleus, via factors that are obtained from \textit{ab initio} atomic structure calculations ~\cite{Safronova:2013, Porsev:2021}. 
The values for the nuclear moments, as calculated from the new experimental hyperfine constants and the recent atomic structure coefficients ~\cite{ Porsev:2021}, are included in Tab.~\ref{hfsAB-alt}. The stated uncertainties combine the contributions from the experimental hfs constants and the \textit{ab initio} calculations. 

For the magnetic moment, the more accurate value obtained here from the 5F$_{5/2}$ level $\mu=0.365(3)\,\mu_N$ agrees well with the value $\mu=0.366(6)\,\mu_N$ from Ref.~\cite{Porsev:2021}. $\mu_N$ is the nuclear magneton. The 6D$_{5/2}$ level is excluded from the final value in Ref.~\cite{Porsev:2021} due to the larger uncertainty of the \textit{ab initio} calculation for this state, which also affects the value in Tab.~\ref{hfsAB-alt}. For the electric quadrupole moment our results for the 5F$_{5/2}$ and the 6D$_{5/2}$ level agree well, and also with the value of $Q/e\,=\,3.11(2)\cdot10^{-28}$\,m$^2$ (with $e$: elementary charge) from Ref.~\cite{Porsev:2021}, obtained from the weighted average of four states in $^{229}$Th$^{3+}$. In both of the investigated levels, the uncertainty in $Q$ is dominated by the contribution from the \textit{ab initio} calculations.

\begin{figure}[b]
\includegraphics[scale=0.325]{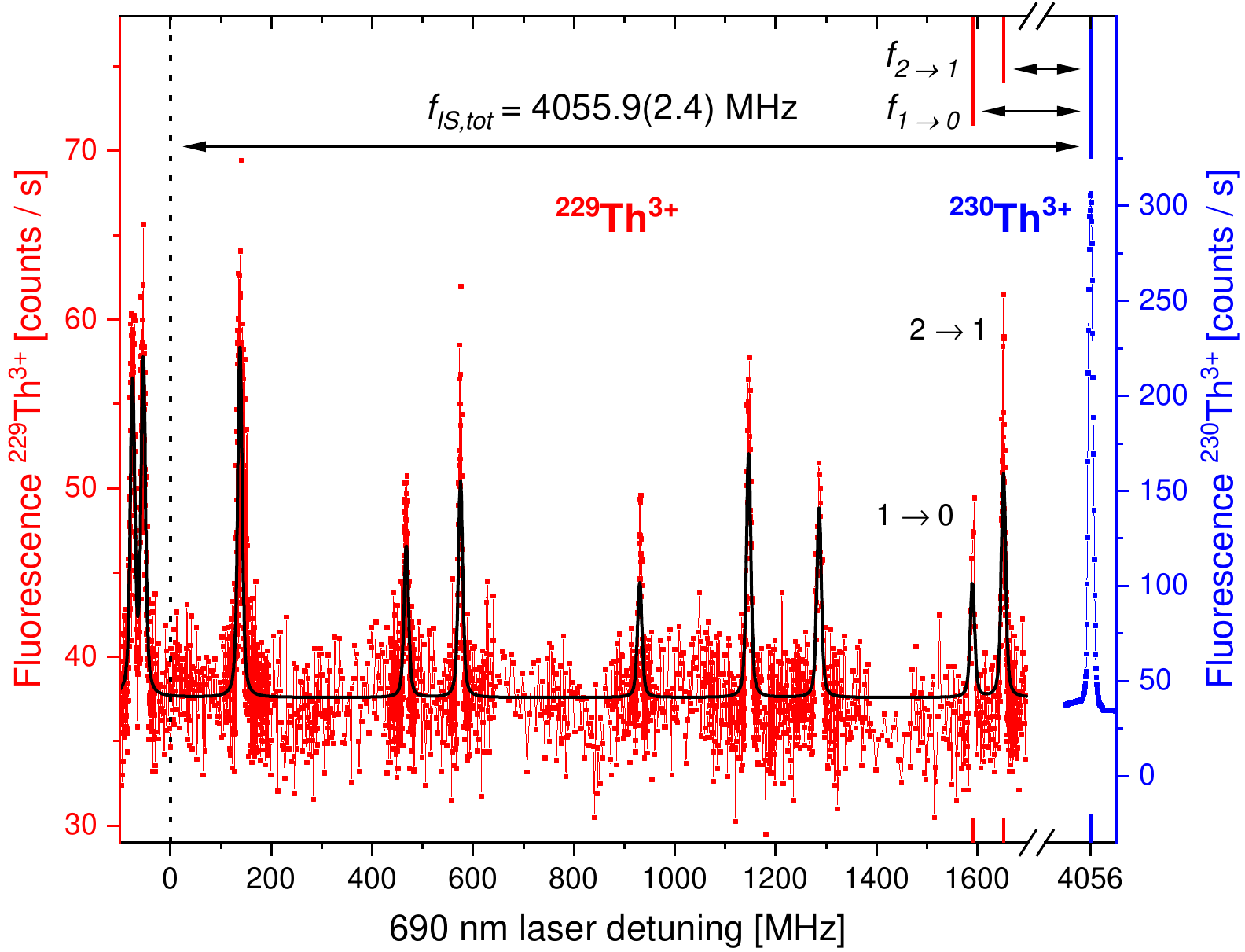}
\caption{\label{fig:Isotopenshiftbild} Isotope shift between $^{229}$Th and $^{230}$Th with the observed fluorescence signals of $^{229}$Th$^{3+}$ (red, left ordinate axis) and $^{230}$Th$^{3+}$ (blue, right ordinate axis) of the 5F$_{5/2}$\,$\rightarrow$\,6D$_{5/2}$ transition. Each data point was averaged over $\approx$\,16 s. To determine the isotope shift the individual frequency intervals between the $^{230}$Th$^{3+}$ resonance and the two outer $^{229}$Th$^{3+}$ hfs components are measured: $f_{2\,\rightarrow\,1}(F=2\,\rightarrow\,F'=1) = 2403.3(8)$\,MHz and $f_{1\,\rightarrow\,0}(F=1\,\rightarrow\,F'\,=\,0) = 2464.8(3)$\,MHz, respectively. The total isotope shift $f_{IS,tot}$ is retrieved by adding the frequency difference of the corresponding transition to the calculated center of the total hfs spectrum.
}
\end{figure}

The experimental system makes it possible to determine the isotope shift between $^{229}$Th$^{3+}$ and $^{230}$Th$^{3+}$ using the rotary vacuum feedthrough to select the isotope that is loaded into the trap.
For $^{230}$Th there is no hfs due to its vanishing nuclear spin (\textit{I} = 0). 
The measurement of the isotope shift is based on the laser frequency difference between the single $^{230}$Th$^{3+}$ resonance and the two $^{229}$Th$^{3+}$ components of the F$_{5/2}$\,$\rightarrow$\,6D$_{5/2}$ (690\,nm) transition at the highest frequencies, $F=1 \rightarrow F'= 0$ and $F = 2 \rightarrow F'=1$. 

For the detection of the $^{230}$Th$^{3+}$ resonance signal similar settings as described in Ref.~\cite{Zitzer:2024} were used. In Fig.~\ref{fig:Isotopenshiftbild} the measured $^{230}$Th$^{3+}$ fluorescence signal and the frequency shift to the $^{229}$Th$^{3+}$ hfs center is presented.
For the two $^{229}$Th$^{3+}$ resonances, an individual shift to the center of $^{230}$Th$^{3+}$ resonance of 2464.8(3)~MHz is measured for the $F=1\rightarrow F'= 0$ component and 2403.3(8)~MHz for the $F =2 \rightarrow F'= 1$ component. The total isotope shift is determined with respect to the center of the $^{229}$Th$^{3+}$ hyperfine structure, that is calculated from equation~(\ref{eqn:hfsenergie}) using the fitted \textit{A} and \textit{B} constants. The result for the shift is $\delta\nu^{229,230}=\nu^{229}-\nu^{230}=- 4055.9(2.4)$\,MHz. The uncertainty is dominated by the uncertainties of the measured hfs \textit{A} and \textit{B} constants. 
Combining the isotope shift and the frequency of 434.28491(7)\,THz for $^{230}$Th$^{3+}$ from the previous measurement~\cite{Zitzer:2024}, the center of the hfs for the 5F$_{5/2}$\,$\rightarrow$\,6D$_{5/2}$ transition in $^{229}$Th$^{3+}$ is located at a frequency of 434.28085(7)\,THz in agreement with the value of 434.280888(31)\,THz from Ref.~\cite{Campbell_Diss:2011}.

At the resolution of the present experiment, the isotope shift is well described as a field shift $\delta\nu^{A,A'}$, neglecting small contributions from the mass differences (few percent of $f_{IS,tot}$; ~\cite{Berengut_Fconst:2009, Safronova:2018}). With this, it can be expressed as the product of the change of the nuclear root-mean-square radius between two isotopes \textit{A} and $A'$, $\delta\langle r^{2}\rangle^{A,A'}$, and the field shift coefficient $F_{fs}$: $\delta\nu^{A,A'} \approxeq F_{fs}\ \delta\langle r^{2}\rangle^{A,A'}$.

So far, the experimental data that has been evaluated for the nuclear radii in the sequence of thorium isotopes~\cite{Angeli:2013} originate from a single laser spectroscopy experiment with trapped Th$^+$ ions~\cite{Kaelber:1989} with uncertainties in the range of 30 - 200\,MHz. With the data presented here and in Ref.~\cite{Campbell:2011} from laser-cooled Th$^{3+}$ ions, it is now possible to combine the two data sets for a test of consistency, similar to a King plot analysis~\cite{King:1963}. For the two isotope pairs and the two transitions, one obtains excellent agreement of the ratios of the shifts and the field shift constants (see Tab.~\ref{isotopeshift}), indicating the consistency of the four experimental data points within their stated uncertainties. Further improvements in the accuracy of Th nuclear radii will therefore predominantly depend on atomic structure theory ~\cite{Porsev:2021, Dzuba_Fconst:2023}. Due to the simpler electronic structure of Th$^{3+}$ in comparison to Th$^+$, the field shift coefficients might be obtainable with lower uncertainty. Presently, the value of the field shift coefficient  for the Th$^{3+}$ transition 5F$_{5/2}$\,$\rightarrow$\,6D$_{5/2}$ of $F_{fs} = 34.5$\,GHz/\,fm$^{2}$ ~\cite{Dzuba_Fconst:2023} is determined on the basis of the shifts of the two individual levels with estimated uncertainties of 10\,\%.

In conclusion, the hyperfine structure constants \textit{A} and \textit{B} for the 5F$_{5/2}$ and 6D$_{5/2}$ levels of $^{229}$Th$^{3+}$ and the isotope shift to $^{230}$Th$^{3+}$ are determined from laser spectroscopy of sympathetically cooled $^{229}$Th$^{3+}$ ions. 
In comparison to previous experimental results \cite{Campbell:2011}, the uncertainties in the hfs constants are reduced by a factor of about three, permitting improved determinations of the $^{229}$Th magnetic dipole and electric quadrupole moment. Future steps with the experiment are targeting the observation of the $^{229m}$Th isomer. Earlier data on the hfs of the isomer and on the isomer shift have been obtained from two transitions in Th$^{2+}$~\cite{Thielking:2018} and, more recently, also from the 1088\,nm line of Th$^{3+}$~\cite{Yamaguchi:2024}. Obtaining more precise information on the charge distribution in the isomer is required for estimating the sensitivity of a $^{229}$Th nuclear clock to the value of the fine structure constant in fundamental tests~\cite{Peik:2021,Beeks:2024_finestructureconstant}. 
Following the experiments on laser excitation of the isomeric state in thorium-doped solids ~\cite{Tiedau:2024,ElwellHudson:2024,Zhang:2024}, achieving this feature with laser-cooled thorium ions and a narrow-linewidth laser at 148\,nm would be a 
significant step towards a highly accurate nuclear optical clock \cite{Peik:2003}.

We would like to thank Thomas Leder, Martin Menzel and Andreas Hoppmann for their expert technical support, and J{\"o}rg Runke and Christoph Mokry for the production of the recoil ion sources. We are grateful to Marianna Safronova, Sergey Porsev and Niels Irwin for helpful discussions. 
This work has been funded by the European Research Council (ERC) under the European Union’s Horizon 2020 research and innovation programme (Grant Agreement No. 856415), the Deutsche Forschungsgemeinschaft (DFG) – SFB 1227 - Project-ID 274200144 (Project B04), TACTICA - Project-ID 495729045 and by the Max-Planck-RIKEN-PTB-Center for Time, Constants and Fundamental Symmetries.

\end{document}